\documentclass[%
reprint,amsmath,amssymb,aps,nofootinbib]{revtex4-1}

\usepackage{comment} 
\usepackage{subcaption,graphicx}% Include figure files
\usepackage{dcolumn}% Align table columns on decimal point
\usepackage{bm}% bold math
\usepackage{amsmath}
\usepackage{slashed}
\usepackage{bm}%
\usepackage[colorlinks=true,linkcolor=blue]{hyperref}%
\expandafter\ifx\csname package@font\endcsname\relax\else
 \expandafter\expandafter
 \expandafter\usepackage
 \expandafter\expandafter
 \expandafter{\csname package@font\endcsname}%
\fi
\usepackage{comment}
\usepackage[normalem]{ulem}
\usepackage{color}

\definecolor{darkcyan}{cmyk}{1,0,0,0.4}

\newcommand{\beq}{\begin{equation}}
\newcommand{\eeq}{\end{equation}}
\newcommand{\bsp}{\begin{split}}
\newcommand{\esp}{\end{split}}
\newcommand{\bit}{\begin{itemize}}
\newcommand{\eit}{\end{itemize}}

\def\ltap{\raisebox{-.4ex}{\rlap{$\sim$}} \raisebox{.4ex}{$<$}} 
\def\gtap{\raisebox{-.4ex}{\rlap{$\sim$}} \raisebox{.4ex}{$>$}}
\def\barr{\begin{array}}
\def\earr{\end{array}}
\def\dis{\displaystyle}

\graphicspath{ {figures/} }

\begin{document}

\title{Anomalous gauge couplings vis-$\grave{a}$-vis  $(g-2)_\mu$ and flavor observables}
\author{Debajyoti Choudhury}
\email{debchou.physics@gmail.com}
\author{Kuldeep Deka}
\email{kuldeepdeka.physics@gmail.com}
\author{Suvam Maharana}
\email{smaharana@physics.du.ac.in}
\author{Lalit Kumar Saini}
\email{sainikrlalit@gmail.com}
\affiliation{Department of Physics \& Astrophysics, University of Delhi}

\preprint{APS/123-QED}


\begin{abstract}

We reassess non-standard triple gauge couplings in the
light of the recent $(g-2)_\mu$ measurement at FNAL, the new lattice
theory result of $(g-2)_\mu$ and the updated measurements of several $B$-decay modes. In the framework of
SMEFT, three bosonic dimension-6 operators are invoked to parametrize
physics beyond the Standard Model and their contributions to such low-energy observables computed. Constraints on the corresponding Wilson coefficients are
then derived from fits to the current experimental bounds on the
observables and compared with the most stringent ones
available from the 13 TeV LHC data in the $W^+ W^-$ and $W^\pm Z$ production channels.

\end{abstract}  

%Use showkeys class option if keyword
\maketitle

%%%%%%%%%%%%%%%%%%%%%%%%%%%%%%%%%%%%%%%%%%%%%%%%%%%%%%%%%%%%%%%%%%%

\section{\label{sec:Introduction} Introduction}

Despite its remarkable compatibility with most experimental
observations in elementary particle physics, the Standard Model (SM)
is plagued by a number of shortcomings warranting explorations in the vistas beyond.  While some of the experimental
``discrepancies'' such as the observation of neutrino masses and
mixings are relatively easy to address, others such as anomalies in
$B$-decays, the seeming deviations in the anomalous magnetic moments
of the muon and the electron or that in the forward-backward
asymmetry in $e^+e^- \to b \bar b$ at the $Z$-peak have not only been
long-standing, but also beg for more complicated solutions.

Solutions to
individual issues are, of course, relatively easy to construct, but a
coherent explanation is much more difficult to achieve, some recent
examples being afforded by efforts that address $(g-2)_\mu$ in
conjunction with discrepancies such as those in low-energy flavour
anomalies \cite{Du:2021zkq,Ban:2021tos,Darme:2021qzw,
  Bhattacharya:2021ggm}, neutrino masses \cite{Nomura:2021oeu,
  Zhang:2021dgl, Borah:2021khc}, dark matter \cite{Arcadi:2021yyr, Bai:2021bau,
  Choudhury:2020xui,Borah:2021jzu} and others \cite{Abdughani:2021pdc, Babu:2021jnu,
  Keung:2021rps, Amaral:2021rzw}. The difficulty of the exercise can be gauged
by the fact that attempted explanations of even a single deviation are
constrained by the need to be in accordance with other observables so as to admit only a relatively small parameter
space~\cite{Choudhury:2017ijp,Altin:2017sxx,Choudhury:2019sxt,Altmannshofer:2020axr,Choudhury:2021nib,Cao:2021tuh,Ke:2021kgy,Kim:2021suj}.

The absence of any new resonances at the LHC strongly suggests that any
new physics (NP) explanation of the extant discrepancies would require
the operative scale $\Lambda$ to be at least a few TeVs or
larger. Thankfully, even such a little hierarchy between
$\Lambda$ and the electroweak scale validates the use of an effective
field theory (EFT) to address these discrepancies
\cite{Falkowski:2016cxu,Choudhury:2020cpm,Buttazzo:2020eyl,Aebischer:2021uvt}.
While a given ultraviolet-complete theory would yield, on the heavy
fields being integrated out, a very specific structure for the ensuing
EFT (in other words, specific relations between the Wilson
coefficients), in the absence of such a theoretically motivated
completion, the coefficients are completely independent.  It has to be
realized, though, that even apparently independent gauge-invariant
operators may be related through equations of motions and the
literature has seen several different choices for the truly
independent operators, {\em viz.}  the Warsaw
basis~\cite{BUCHMULLER1986621,Grzadkowski:2010es}, the HISZ
basis~\cite{PhysRevD.48.2182}, the SILH basis~\cite{Giudice:2007fh}
etc. The lack of discernible deviations from the SM expectations have
led to constraints on such SMEFT operators, whether of the 4-fermion
form~\cite{Falkowski:2017pss} or otherwise
~\cite{Faham:2021zet,Aebischer:2021uvt,Bellan:2021dcy,Yin:2021mls,Ethier:2021bye,ATLAS:2020yat}. In
this work, we carry on in this spirit, sticking to the HISZ basis.

Of much recent interest has been the Fermi National Accelerator
Laboratory (FNAL) measurement~\cite{Muong-2:2021ojo} of the anomalous
magnetic moment of the muon, namely $a_\mu \equiv [(g-2)/2]_\mu$. The
result is consistent with the previous one from the Brookhaven
National Laboratory (BNL) measurement, and it is customary to combine
the two so as to reduce the experimental errors.  As is well-known,
$a_\mu$ receives many corrections within the SM, of which a
particularly difficult calculation is that for the contribution from
the hadronic vacuum polarization. The traditional method has been to
use dispersion relations alongwith experimental results and if this is
adopted, the experimental result would imply a discrepancy $\Delta
a_\mu^{\mbox{\tiny DISP}} = 251(59) \times 10^{-11}$, a $4.2 \sigma$
deviation from the the SM \cite{Aoyama:2020ynm,Muong-2:2021ojo,Davier:2017zfy,*Keshavarzi:2018mgv,*Colangelo:2018mtw,*Hoferichter:2019mqg,*Davier:2019can,*Keshavarzi:2019abf,*Keshavarzi:2019abf,*Kurz:2014wya,*Chakraborty:2017tqp,*Borsanyi:2017zdw,*Blum:2018mom,*Giusti:2019xct,*Shintani:2019wai,*FermilabLattice:2019ugu,*Gerardin:2019rua,*Aubin:2019usy,*Giusti:2019hkz,*Melnikov:2003xd,*Masjuan:2017tvw,*Colangelo:2017fiz,*Hoferichter:2018kwz,*Gerardin:2019vio,*Bijnens:2019ghy,*Colangelo:2019uex,*Pauk:2014rta,*Danilkin:2016hnh,*Jegerlehner:2017gek,*Knecht:2018sci,*Eichmann:2019bqf,*Roig:2019reh,*Colangelo:2014qya,*Blum:2019ugy,*Melnikov:2003xd,*Masjuan:2017tvw,*Colangelo:2017fiz,*Hoferichter:2018kwz,*Gerardin:2019vio,*Bijnens:2019ghy,*Colangelo:2019uex,*Pauk:2014rta,*Danilkin:2016hnh,*Jegerlehner:2017gek,*Knecht:2018sci,*Eichmann:2019bqf,*Roig:2019reh,*Blum:2019ugy,*Aoyama:2012wk,*Aoyama:2019ryr,*Czarnecki:2002nt,*Gnendiger:2013pva,*Davier:2017zfy,*Keshavarzi:2018mgv,*Colangelo:2018mtw,*Hoferichter:2019mqg,*Davier:2019can,*Keshavarzi:2019abf,*Kurz:2014wya,*Melnikov:2003xd,*Masjuan:2017tvw,*Colangelo:2017fiz,*Hoferichter:2018kwz,*Gerardin:2019vio,*Bijnens:2019ghy,*Colangelo:2019uex,*Pauk:2014rta,*Danilkin:2016hnh,*Jegerlehner:2017gek,*Knecht:2018sci,*Eichmann:2019bqf,*Roig:2019reh,*Blum:2019ugy,*Colangelo:2014qya,*Aoyama:2012wk,*Aoyama:2019ryr,*Czarnecki:2002nt,*Gnendiger:2013pva,*Davier:2017zfy,*Keshavarzi:2018mgv,*Colangelo:2018mtw,*Hoferichter:2019mqg,*Davier:2019can,*Keshavarzi:2019abf,*Kurz:2014wya,*Melnikov:2003xd,*Masjuan:2017tvw,*Colangelo:2017fiz,*Hoferichter:2018kwz,*Gerardin:2019vio,*Bijnens:2019ghy,*Colangelo:2019uex,*Blum:2019ugy,*Colangelo:2014qya}.  On the other hand, the adoption of the
Lattice QCD results from the Budapest-Marseille-Wuppertal (BMW)
collaboration \cite{Borsanyi:2020mff} for the same significantly
reduces the deviation down to $\Delta a_\mu^{\mbox{\tiny BMW}} = 107(69) \times
10^{-11}$ or an agreement at the $2\sigma$ level\footnote{
    It should be recognized, though, that while such
    a positive change in the
      hadronic vacuum polarization (HVP) minimizes the apparent
      $(g-2)_\mu$ discrepancy, it could possibly engender conflicts with the
      global EW fit prediction of the hadronic contributions to the
      QED coupling $\Delta \alpha^{(5)}_{\mbox{had}}$ or elsewhere in
      low-energy hadron phenomenology~\cite{Crivellin:2020zul,Keshavarzi:2020bfy,Colangelo:2020lcg}.} (see ref
\cite{Keshavarzi:2021eqa,Gerardin:2020gpp, Keshavarzi:2019xgi} for
reviews).

Also of interest are $B$ physics observables, where neutral current $b
\rightarrow s l l $ transitions have been showing persistent
discrepancy from the SM values in recent years with the most recent
result being of the $R_K$ anomaly and $BR(B_s \rightarrow \mu^+
\mu^-)$. In Moriond 2021, the LHCb collaboration has reported the
measured value of $R_K$~\cite{LHCb:2021trn} to be
$0.846^{+0.044}_{-0.041}$ in the 1.0 GeV $\leq$ $q^2$ $\leq$ 6.0
GeV$^2$ bin. With the central value remaining virtually unchanged from
the earlier result~\cite{LHCb:2019hip}, and the errors shrinking by
almost 30\%, this has strengthened the deviation from the erstwhile
$2.5\sigma$ level to $3.1\sigma$. Most interestingly, the value of
$BR(B_s \rightarrow \mu^+ \mu^-)$ as reported by LHCb is
$3.09^{+0.483}_{-0.443} \times 10^{-9}$~\cite{LHCb:2021awg} and
compatible
with the SM prediction within $1 \sigma$ whereas the previous world
average (ATLAS, CMS and LHCb) of $2.69^{+0.37}_{-0.35} \times 10^{-9}$
was below the SM prediction by $2 \sigma$.

In the present paper, we reexamine possible anomalous
self-interactions of the electroweak gauge bosons in the light of
these experiments and other such.  Concentrating on three particular
dimension-6 terms in the SMEFT Lagrangian that lead to anomalous
triple gauge boson couplings (TGCs), we evaluate the corresponding
one-loop contributions to both $(g-2)_\mu$ and $(g-2)_e$, another
observable that shows a discrepancy, albeit smaller as well as certain
electroweak precision measurements.  While similar exercises have been
undertaken in the past~\cite{Burgess:1993qk,Arbuzov:2021lob}, our
study differs in our inclusion of not only the direct constraints from
CMS and ATLAS~\cite{CMS:2019ppl,CMS:2020mxy,ATLAS:2019rob}, but also
the recent results from the LHCb experiment.  Assuming that the
aforementioned operators are the leading ones, we show that radiative
and rare $B$ and $K$ decays such as $B \to X_s \gamma$, $B_s \to \mu^+
\mu^-$, $B \to X_s \ell^+ \ell^-$, $B \to K^{(\ast)} \mu^+ \mu^-$,
$B_s \to \phi \mu^+ \mu^-$, $K \to \pi \nu \bar \nu$ provide very
important constraints. Even though a comprehensive study with the
preceding assumption seems quite restrictive at the outset and would,
presumably, have very little to say about the mentioned anomalies in
terms of a unified explanation, we find that the ensuing results have
interesting implications nonetheless, especially
with regards to constraining
certain classes of new physics models that could potentially serve to
explain such discrepancies more efficiently.

This paper is organised as follows. In section II we will introduce
the effective Lagrangian for the Electro-Weak (EW) Gauge Bosons and
relate the couplings to the Wilson coefficients of the relevant
dimension-6 SMEFT operators. In Section III we discuss the anomalous
contribution to $(g-2)_\mu$, $b \rightarrow s \gamma$, $b \rightarrow
s \mu^+ \mu^-$ and $Z \rightarrow b \bar{b}$ using the effective
Lagrangian. In Section IV we will present our results using the
current experimental observations and also discuss future projections
in context of $(g-2)$ experiments. Finally we conclude in Section V.

\section{\label{sec:Model} Effective Lagrangian for the Gauge Sector}

With $\Lambda$ as the characteristic scale of the UV-complete theory,
the EFT operative between the electroweak scale and $\Lambda$ would be
described by a Lagrangian as an expansion in $\Lambda^{-1}$ with each
term being invariant under the full SM gauge group. As the only possible
dimension-5 terms do not respect global lepton
number, we are eschewing this, and to the leading order, the effective
Lagrangian can be expressed as
\begin{eqnarray}
	\mathcal{L} &=& \mathcal{L}_{SM} + \sum_i\, \frac{c_i}{\Lambda^2}\, \mathcal{O}_i 
\end{eqnarray}
where $\mathcal{L}_{SM}$ is the SM lagrangian and $\mathcal{O}_i$ is
the set of gauge-invariant dimension-6 SMEFT operators with
corresponding Wilson coefficients $c_i$. We make a further assumption
that the superheavy fields couple primarily to the bosonic sector and
thus the leading corrections are those that involve the latter rather
than the SM fermions\footnote{This, of course, is a strong assumption
  and we would return to this point later.}. The assumption of larger
Wilson coefficients for the bosonic operators than those for the
4-fermion operators (despite both sets nominally being of
mass-dimension six) may seem an {\em ad hoc} prescription. However,
there exist a plethora of scenarios wherein this could (and is indeed
likely to) emerge naturally.  Perhaps the most famous of these are
Randall-Sundrum-like scenarios with bulk fermions and bosons. The
localizations of the light fermions as dictated by the warping,
whether a single one~\cite{Chang:1999nh, Csaki:2002gy, Agashe:2003zs}
or a multiple and nested one~\cite{Arun:2015kva, Arun:2016ela},
ensures that the overlap integrals for the KK-gauge bosons with the SM
fermions are much smaller than those with the SM bosons. This,
immediately, leads to an hierarchy in the Wilson coefficients as
examined in this analysis\footnote{Such a hierarchy would obviously be
  manifested when the bosonic operators can contribute to an
  observable at the Born level itself. On the other hand, even when
  they can contribute only at the loop-level, the hierarchy---itself
  depending on the features of the theory such as the extent of
  warping or the profiles of the low energy fields---can be such that
  these still overcome the pure-fermionic or mixed fermion-boson
  operators. We shall implicitly assume this to be so.}.  Indeed, any
model that would include additional particles that couple
preferentially to the electroweak bosons rather than to the SM
fermions could, in principle, result in larger $c_i$ corresponding to
bosonic operators rather than four-fermion operators involving the SM
fields alone. Examples are provided by numerous scenarios that contain
higher gauge-multiplet fermions such as those in a wide class of
scenarios seeking to explain, amongst others, neutrino masses and
mixings
\cite{Babu:2009aq,*Bonnet:2009ej,*Li:2009mw,*Picek:2009is,*Liao:2010ku,*Delgado:2011iz,*Kumericki:2011hf,*Bonnet:2012kz,*Kumericki:2012bh,*Ma:2013tda,*McDonald:2013hsa,*Ma:2014zda,*Yu:2015pwa,*Ko:2015uma,*Cepedello:2017lyo,*Anamiati:2018cuq,*KumarAgarwalla:2018nrn,*Agarwalla:2018xpc,*Arbelaez:2019cmj,*Kumar:2019tat,*Ashanujjaman:2020tuv}.
In this sense, the analysis presented here serves to assess the
viability of such explicit models (many of which, potentially, lead to
interesting, albeit complicated signals at the
LHC~\cite{Kumar:2021umc,Ashanujjaman:2022cso}) when juxtaposed with
the most recent limits on the low-energy observables, to be described
later.

The purely bosonic interactions are expressible in terms of 11
operators\cite{PhysRevD.48.2182}, of which two
($\mathcal{O}_{\Phi, 2}, \mathcal{O}_{\Phi, 3}$) contribute, at the tree-level, solely to the Higgs
self-interaction.  Containing terms
  proportional to the gauge boson kinetic terms, $\mathcal{O}_{BB}$
and $\mathcal{O}_{WW}$ not only lead to a finite
renormalization of the $W$ and $B$ fields respectively,
but also contribute to the $H \rightarrow
\gamma\gamma/ Z\gamma$ decays, and, hence, are constrained by
  these.  Similarly,
$\mathcal{O}_{\Phi, 1}$ contributes to the $Z$ boson mass but not to
the $W$ mass and hence leads to deviations of the $\rho$ parameter
from 1. Finally, $\mathcal{O}_{DW}$ and $\mathcal{O}_{DB}$ lead to an
anomalous running of the QED fine structure constant and of the weak
mixing angle.

Concentrating, for the sake of simplicity, on only those operators
that would leave the largest imprint on the observables of interest, we
would examine~\cite{PhysRevD.48.2182}
\begin{equation} \label{eqn:EFT}
\begin{split}
\mathcal{O}_{WWW}\ &=\ \mbox{Tr}\left[\hat{W}_{\mu}^{ \nu}\hat{W}_{\nu}^{ \rho}\hat{W}^{\mu}_{\rho}\right] \\
\mathcal{O}_{W}\ &=\ \left( D_\mu \Phi\right)^{\dagger}\hat{W}^{\mu \nu}\left( D_\nu \Phi\right) \\
\mathcal{O}_{B}\ &=\ \left( D_\mu \Phi\right)^{\dagger}\hat{B}^{\mu \nu}\left( D_\nu \Phi\right),
\end{split}
\end{equation}
assuming the others to be absent (or, at the least, severely constrained as discussed above). Here, $ D_\mu \Phi=
\left(\partial_\mu +ig\frac{\sigma^a}{2}W_{\mu}^a+i\frac{g'}{2} B_\mu
\right)\Phi$, $\hat{W}_{\mu \nu}=ig\frac{\sigma^a}{2}W_{\mu \nu}^a$
and $\hat{B}_{\mu \nu}=i\frac{g'}{2}B_{\mu \nu}$. Note that, while
analogous $CP$-odd operators (obtained by replacing a field-strength
tensor by its dual) would exist as well, these are irrelevant for our
analysis as we deal with only $CP$-even observables\footnote{ The leading contributions of the $CP$-odd
  dimension-6 operators to the $CP$-even observables would appear only
  at second order in the corresponding Wilson coefficients. In the
  ensuing analysis, however, we limit contributions only upto a linear
  order in the WCs. Assuming higher order terms would further require
  the inclusion of operators of similar mass-dimensions (dimension-8
  and higher) in the theory which is beyond the scope our current
  study.}.

The operators in eqn.(\ref{eqn:EFT}) give rise to, amongst other terms,
anomalous triple gauge
boson couplings (TGCs). The latter are also often parametrized
in terms of a convenient
phenomenological Lagrangian~\cite{PhysRevD.48.2182, Gounaris:1996rz} namely
\begin{equation} \label{eqn:LEP}
\begin{split}
\mathcal{L}^{WWV}_{eff}=g_{WWV}\Big\{ & g_1^V \left( \tilde W^{-}_{\mu \nu} \tilde W^{+ \nu}-\tilde W^{+}_{\mu \nu} \tilde W^{- \nu} \right) V^\mu \\
&+\kappa_V \tilde W^{+}_{\mu} \tilde W^{-}_{\nu} \tilde V^{\mu \nu}\\
 &
+ \frac{\lambda_V}{m_W^2}\tilde W_{\mu}^{+ \nu} \tilde W_{\nu}^{- \rho}  {\tilde{V}_{\rho}}^{\, \,\mu}
\Big\},
\end{split} 
\end{equation}
with $V\equiv \gamma, Z$. Here, $g_{WW\gamma}=e$, $g_{WWZ}=e \cot \theta$ (with $\theta$ being
  the Weinberg angle) and the field strengths correspond to only the
  abelian part,  $\tilde W_{\mu \nu}= \partial_\mu W_\nu - \partial_\nu
W_\mu$ and $\tilde V_{\mu \nu}= \partial_\mu V_\nu - \partial_\nu
V_\mu$. Within the SM, we have $g_1^V=\kappa_V=1$, $\Delta g_1^\gamma$ = 0 and
$\lambda_V=0$. In other words, $\Delta \kappa_V\equiv\kappa_V-1$,
$\Delta g_1^Z \equiv g_1^Z-1$ and $\lambda_V$ suitably define the
anomalous couplings, and, post symmetry-breaking, can be related to
the Wilson coefficients $c_W$, $c_B$ and $c_{WWW}$ as follows:
\begin{equation}\label{eqn:transf}
\begin{split}
\Delta g_1^Z&=c_W \frac{m_Z^2}{2\Lambda^2} \\
\Delta \kappa_Z &= \left[c_W-s^2_\theta \left( c_W + c_B \right)\right]\frac{m_Z^2}{2\Lambda^2} \\
\Delta \kappa_\gamma &= \left(c_W + c_B \right)\frac{m_W^2}{2\Lambda^2} \\
\lambda_\gamma&=\lambda_Z=\frac{3 m_W^2 g^2}{2 \Lambda^2}c_{WWW}.
\end{split}
\end{equation}
In the above and in the following sections we use the notation $s_\theta=\sin{\theta}$ and $c_\theta=\cos{\theta}$.

Although the operators $\mathcal{O}_{DW}$ and $\mathcal{O}_{BW}$ too
contribute to the anomalous triple gauge couplings, their primary
effect on low-energy physics accrue through the modification of the
gauge boson propagators and, hence, these are not as visible in processes such as
$W^+ W^-$ production in LEP\cite{DERUJULA19923} and LHC experiments.

\section{Contributions to various observables} \label{sec:allobs}

At this stage, the choice of the gauge is an important one. While full
$SU(2) \otimes U(1)$ gauge invariance is manifest in the formulation
of eqn.(\ref{eqn:EFT}), it is not so for the case where
eqn.(\ref{eqn:LEP}) is supposed to encapsulate the entire non-SM part
of the effective Lagrangian. In view of this, and for the sake of
convenience, we adopt the unitary gauge for all our calculations.

\subsection{\label{sec:magmoment}{\boldmath $a_\mu$}}

\begin{figure}[htb]
	\centering
\hspace{0.5cm}
%  \begin{subfigure}{0.2\textwidth}
%      \centering
	\includegraphics[scale=0.08,keepaspectratio=true]{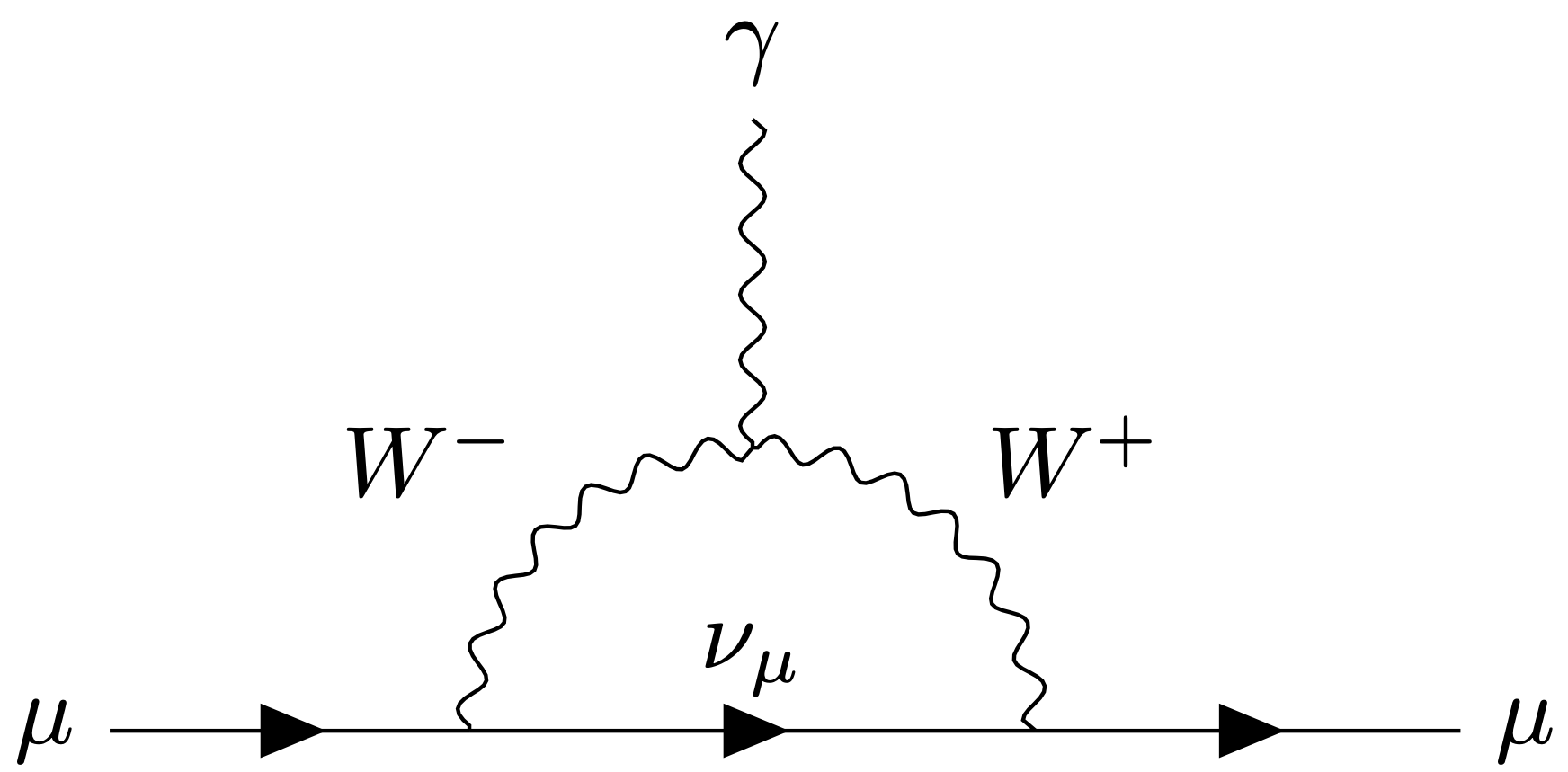}
%      \caption{}	\label{fig:D2}
%  \end{subfigure}%
	\caption{\textit{ Diagram contributing to $\Delta
            a_\mu^{\rm anom}$ at 1-loop involving an aTGC vertex.}}
        \label{fig:muon}
\end{figure}

Starting with the effective Lagrangian as described in the preceding section,
we may now compute the contribution to the magnetic moment of the muon.
At the one-loop order,
the only contributing diagram is
as shown in Fig. \ref{fig:muon}.
To this order, then, the only relevant part of the higher-dimensional
terms is that subsumed in eqn.(\ref{eqn:LEP}), and more specifically
\begin{equation}
\mathcal{L}_{\Delta a_\mu} \supset \frac{g}{\sqrt{2}}\left(\bar{\mu}\gamma^\mu P_L \nu_\mu \right) W^{-}_\mu + \mbox{h.c.} + \mathcal{L}^{WW\gamma}_{\rm eff},
\end{equation}
with $P_L$ being the left-handed projection operator.  Since we
are dealing with a nonrenormalizable Lagrangian, we expect the
integral to be a divergent one (a simple power counting ensuring that
it is only a logarithmic divergence), necessitating the introduction
of a cutoff. A natural choice for the latter is $\Lambda$
itself\footnote{While this identification might seem an {\em ad hoc}
  one, note that the cutoff has to be $\ltap \Lambda$, and given the
  logarithmic nature of the dependence, a small variation would be
  numerically insignificant.}, leading to
\begin{equation} \label{eqn:delamu}
\begin{split}
\Delta a_\mu^{anom.}=\frac{e^2}{48 \pi^2 s^2_\theta}& \frac{m_\mu^2}{m_W^2}\Bigg\{ \Delta \kappa_\gamma \left( \frac{1}{3}+\ln\left[ \frac{\Lambda^2}{m_W^2}\right]\right)\\
& + \lambda_\gamma \left( \frac{7}{6}-\ln\left[ \frac{\Lambda^2}{m_W^2}\right]\right) \Bigg\}.
\end{split}
\end{equation} 
where we have retained only terms at the leading order in
$m_\mu^2/m_W^2$ and neglected $m_\nu$ altogether. This expression can,
of course, be trivially translated in terms of $c_{WWW}$ and the
combination $(c_W + c_B)$ using eqns.(\ref{eqn:transf}).  While
Fig.\ref{fig:muon} represents the only one-loop contribution to
$\Delta a_\mu$ emanating from the Lagrangian of eqn.(\ref{eqn:LEP}),
when the SMEFT language of eqn.(\ref{eqn:EFT}) is used instead,
additional contributions arise from diagrams with a Higgs in the
loop. Contrary to naive expectations, these are {\em not} small
compared to that in eqn.(\ref{eqn:delamu}). The analytic expressions
thereof are more unwieldy though, and, for the sake of breivity we
omit presesnting those. However, we do include such contributions in
our numerical analysis.  Note that the four-point (or higher) vertices
that eqn.(\ref{eqn:EFT}) engenders---and not included in
eqn.(\ref{eqn:LEP})---contribute to $a_\mu$ only at the two-loop or
higher orders.

In the preceding and subsequent loop calculations we adopt dimensional
regularization alongwith the $\overline{\mbox{MS}}$ renormalization
scheme.  While it is often argued that the simple pole at $d = 4$ ($d$
being the number of dimensions) could be straightforwardly exchanged
for a logarithmic dependence on the cutoff $\Lambda$, the said
dependence is better understood as a consequence of the
renormalization group (RG) evolution of the effective operators from
the scale $\Lambda$ down to the EW scale
\cite{PhysRevLett.69.3428}. Note that we ignore the (subdominant)
contributions of terms with higher powers of the logarithm and the EW
couplings which appear in perturbative solutions of the RGEs.

It might seem that the contribution $\mathcal{O}_{WWW}$
is in disagreement with the
results of ref.\cite{Alonso:2013hga}, which were expressed in 
terms of the dipole operator. Of particular significance
is the presence of the 
$\ln (\Lambda^2/m_W^2)$ dependence in our result {\em vis \'a vis} the
apparent absence of $c_{WWW}$ in the anomalous dimension (as computed
in the Warsaw
basis) of $c_{e\gamma}$ (or, equivalently, $c_{eB}$ and $c_{eW}$).
However, a comparison between operators in different bases needs to be
made with care, especially since the dipole operators ({\em sans} the Higgs field) are admissible only post electroweak symmetry breaking. A careful matching
reintroduces this effect, and, indeed our result for $\Delta a_\mu$ is,
consistent with that given in
ref.\cite{Aebischer:2021uvt} obtained from the one-loop matching between
LEFT and SMEFT operators at the electroweak scale.

%%%%%%%%%%%%%%%%%%%%%%%%%%%%%%%%%%%%%%%%%%%%%%%%%%
%\begin{figure}[h!]
%\includegraphics[scale=.16]{rge_plots_new/case1_c_e_gamma}
%\includegraphics[scale=.16]{rge_plots_new/case1_C_HWB}
%\caption{Case 1}
%\end{figure}
%\begin{figure}[h!]
%\includegraphics[scale=.16]{rge_plots_new/case2_c_e_gamma}
%\includegraphics[scale=.16]{rge_plots_new/case2_C_HWB}
%\caption{Case 2}
%\end{figure}
%\begin{figure}[h!]
%\includegraphics[scale=.16]{rge_plots_new/case3_c_e_gamma}
%\includegraphics[scale=.16]{rge_plots_new/case3_C_HWB}
%\caption{Case 3}
%\end{figure}
%%%%%%%%%%%%%%%%%%%%%%%%%%%%%%%%%%%%%%%%%%%%%%%%%%

While we have argued for the Wilson coefficient $c_{e \gamma}$ being,
at best, tiny at the scale $\Lambda$, it is worthwhile to examine
whether it could be amplified at the EW scale as a result of mixing,
under RG evolution, with the other operators in the theory. Note that
the operators we use are a part of the HISZ set
which, in consisting of only eleven purely bosonic operators, does not
constitute a closed set under RG evolution.  However, confined to this
subset, these modulo
a difference in the normalizations, can be exactly mapped onto
  operators in the SILH basis. Inspired by the latter, if we were to
augment the set by the inclusion of the dipole operator ($c_{e
  \gamma}$), its evolution would depend mainly on $\mathcal{O}_{BW}$
($\equiv H^\dagger H W_{\mu \nu} B^{\mu \nu}$) apart from
$\mathcal{O}_{B}$ and $\mathcal{O}_{W}$.  The dependence of $\Delta
a_\mu$ on the last two operators has already been calculated above to
one-loop order and any residual dependence would be further
loop-suppressed and negligible.  The Wilson coefficient for
$\mathcal{O}_{BW} $, on the other hand, is constrained by the LEP
experiments (from oblique corrections) to be tiny
\cite{PhysRevD.48.2182}.  Consequently, any enhancement,
  on evolving down to the EW scale (or even lower) in such an
  explicitly introduced $c_{e\gamma}/\Lambda^2$ is tiny. For example, a value
of starting with $c_{e \gamma}/\Lambda^2 \sim {\cal O}(10^{-3}~ \mbox{TeV}^{-2})$ and $c_{BW}/\Lambda^2 \sim {\cal O}(0.1
~\mbox{TeV}^{-2})$ at $\Lambda=1$ TeV,   the RG evolution down to the EW scale results in a change of less than $1\%$ in either. In other words,
the running effects
  can be neglected.

It is easy to see that an expression analogous to that in
eqn.(\ref{eqn:delamu}) would hold for $\Delta a_e$ with $m_\mu$
replaced by $m_e$. Most importantly, the sign of the new contributions
would be identical in the two cases, which is in disagreement with the
experimental results.

%%%%%%%%%%%%%%%%%%%%%%%%%%%%%%%%%%%%%%%%%%%%%%%%%%%%%%%%%%%%%%%%%%%%
\subsection{\label{sec:flavor_obs} Flavor Observables}

\begin{figure}[ht]
\includegraphics[scale=.08]{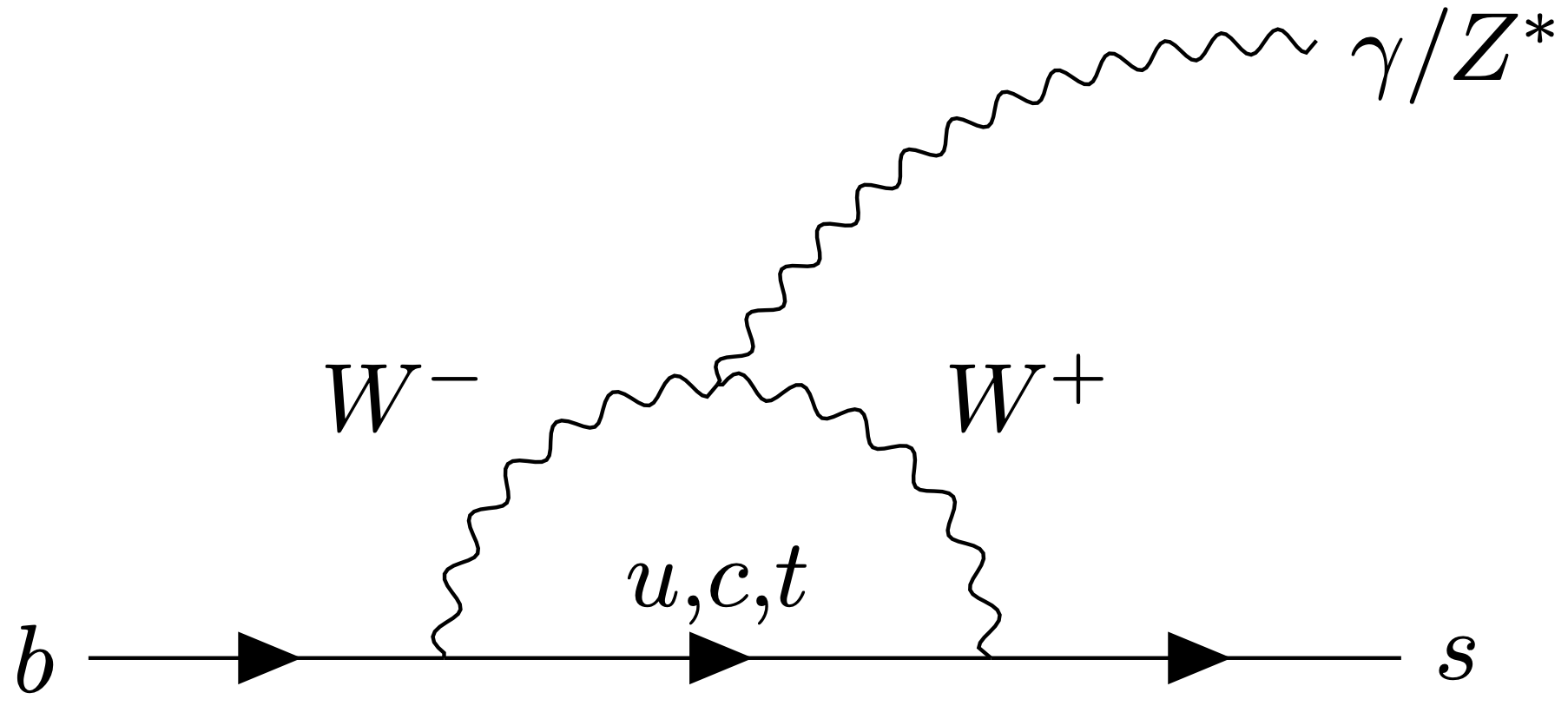}
\caption{\textit{Diagram contributing to $b \rightarrow s\, \gamma(Z^*)$ at 1-loop.}}
\label{fig:bsg}
\end{figure}

The anomalous gauge couplings can also contribute to various
loop-mediated flavour changing neutral current hadronic decays. The said 
decays may occur through a multitude of effective
operators such as the electromagnetic dipole or  semi-leptonic vector and
axial-vector ones, namely

\begin{equation} \label{flavor_operators}
\begin{split}
\mathcal{Q}_{7} &=\frac{e}{(4 \pi)^2} m_b \left(\bar{s}_L\sigma_{\alpha \beta} b_R\right) F^{\alpha \beta}  \\
\mathcal{Q}_{9}&= \frac{e^2}{(4 \pi)^2}\left(\bar{s}_L \gamma_\alpha b_L) (\bar{l}\gamma^\alpha l\right) \\
\mathcal{Q}_{10}&= \frac{e^2}{(4 \pi)^2}\left(\bar{s}_L \gamma_\alpha b_L) (\bar{l}\gamma^\alpha \gamma_5 l\right) \\
\end{split}
\end{equation}
where $L$ and $R$ denote the chirality of the fermionic fields,
$\sigma_{\alpha \beta}=i[\gamma_\alpha,\gamma_\beta]/2$ and $F^{\alpha
  \beta}$ is the electromagnetic field tensor. The $\Delta B,\Delta S =1$
operator is traditionally written as
\begin{eqnarray}
\mathcal{L} &=& \dfrac{4 G_F}{\sqrt{2}} \Big(C_7 Q_7 + C_9 Q_9 + C_{10} Q_{10}\Big)  + H.c
\end{eqnarray} 
with $C_7$, $C_9$ and $C_{10}$ being the corresponding Wilson
coefficients that factorise the short distance physics.  We have
omitted above the right-handed analogues of $C_{9,10}$ as, to the
leading order, the operators of interest do not contribute to these.
The anomalous contributions appear as a result of the diagrams
in Figure~\ref{fig:bsg}.

Although we use dimensional regularisation, it is instructive to trace
the divergences of the 1-loop contributions in generality. The
photonic diagram thus generated mirrors that for $\Delta a_\mu^{\rm
  anom}$ and is only logarithmically divergent. For the $Z$ vertex, on
the other hand, if a momentum cutoff were used instead, the anomalous
contribution from an individual quark loop would be found to be
quadratically divergent (in dimensional regularisation, this is
manifested as a pole at $d=2$).  However, thanks to the GIM mechanism,
the quadratically divergent pieces (as with any other term independent
of the internal quark mass) cancel, leaving behind only a logarithmic
divergence. Of these, the top-quark contribution dominates
overwhelmingly~\cite{Bobeth:2015zqa} and we can fairly approximate ($x
\equiv m_t^2/m_W^2$)
%%%
\beq
C_i \approx (C_i)_{SM} + \left[V_{tb} V_{ts}^* 
 \, \frac{m_W^2}{\Lambda^2} \,\right]
\Delta C_i 
\eeq
with 
\begin{equation}
 \label{eq:DeltaC710calculation}
\barr{rcl}
\Delta C_7 &= & \dis
\frac{- (c_B+c_W) }{8 \left (x-1 \right )^2}
\left[2 x +  \frac{x^3-3 x^2}{\left (x-1 \right )} \, \ln x \right] \\[2.5ex]
&+&\dis \frac{3 g^2 \,c_{WWW}}{8} \left[\hspace{0.25mm} \frac{x^2+x}{\left (x-1 \right )^2}-\frac{2 x^2 \, \ln x }{\left (x-1 \right )^3} \right] \ ,
\\[3.5ex]
\Delta C_9  & = & \dis
\Bigg\{\frac{-(c_B+c_W) }{8 } \,   x\, + \frac{3 c_W}{16} \, \frac{1-4 s^2_\theta}{s^2_\theta} \, x   \Bigg\} \ln \left( \frac{\Lambda^2}{m_W^2} \right) \\[2ex]
&+&\dis \frac{3 g^2 c_{WWW}}{4\, \left (x-1 \right )^2}
 \left[ x - 3 x^2 +\frac{2 x^3 \ln x}{\left (x-1 \right)} \right], \\[3.5ex]
\Delta C_{10}  & =&\dis \dfrac{-3 c_{W}}{16 s^2_\theta} x \ln \left( \frac{\Lambda^2}{m_W^2} \right)\ .
\earr
\end{equation}
The subdominant contributions from the up- and charm-loop can be
easily read off. For $b\rightarrow d$ and $s \rightarrow d$
transitions, analogous analyses follow, but with the identity of the
dominant loop changing.

%%%%%%%%%%%%%%%%%%%%%%%%%%%%%%%%%%%%%%%%%%%%%%%%%%%%%%%%%%%%%%%%%%%%
\subsection{ \label{sec:flavor_zbb} $Z \to b \bar{b}$}

\begin{figure}[htb]
	\centering
\hspace{-2.5cm}
  \begin{subfigure}{0.2\textwidth}
%      \centering
	\includegraphics[scale=0.08,keepaspectratio=true]{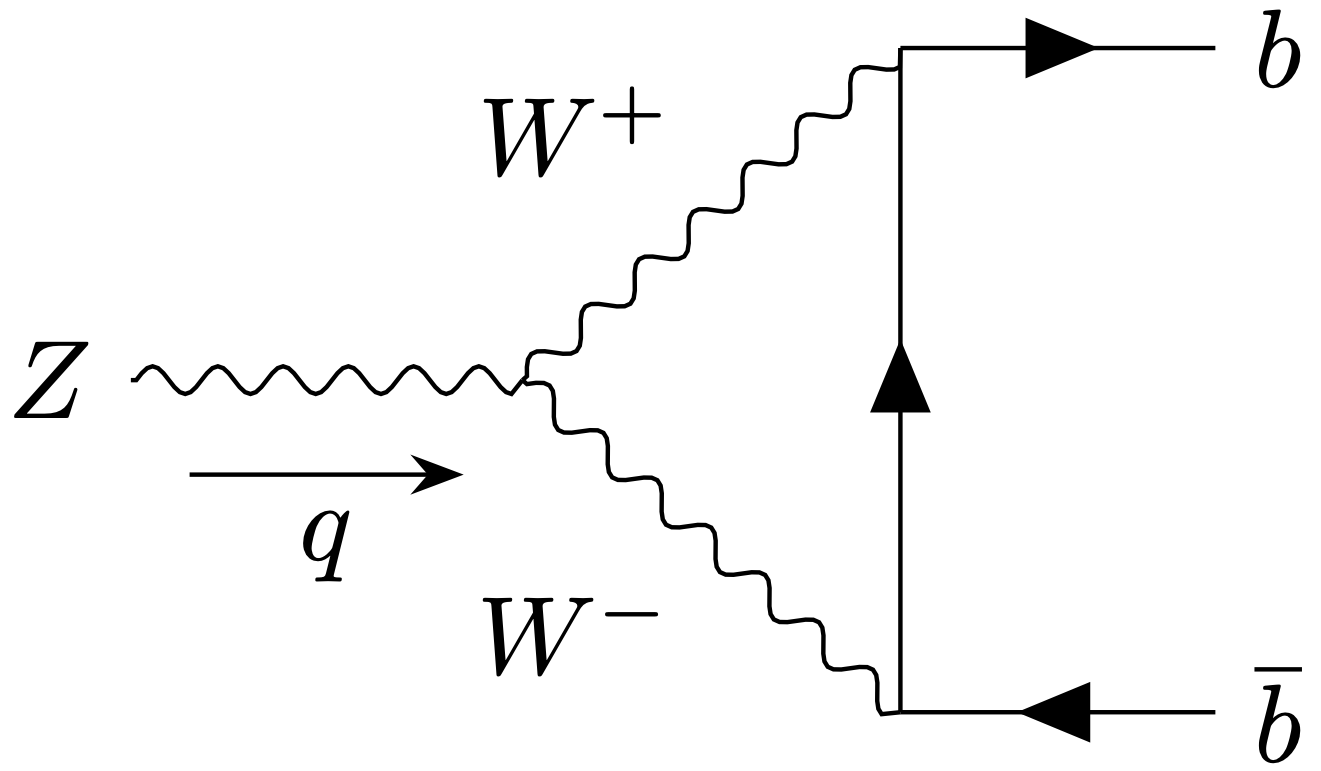}
%\hspace{3.5cm}         
      \caption{}	\label{fig:Z2bbV}
  \end{subfigure}%
  \\
  	\hspace{-3.5cm}
  \begin{subfigure}{0.2\textwidth}
%      \centering
	\includegraphics[scale=0.08,keepaspectratio=true]{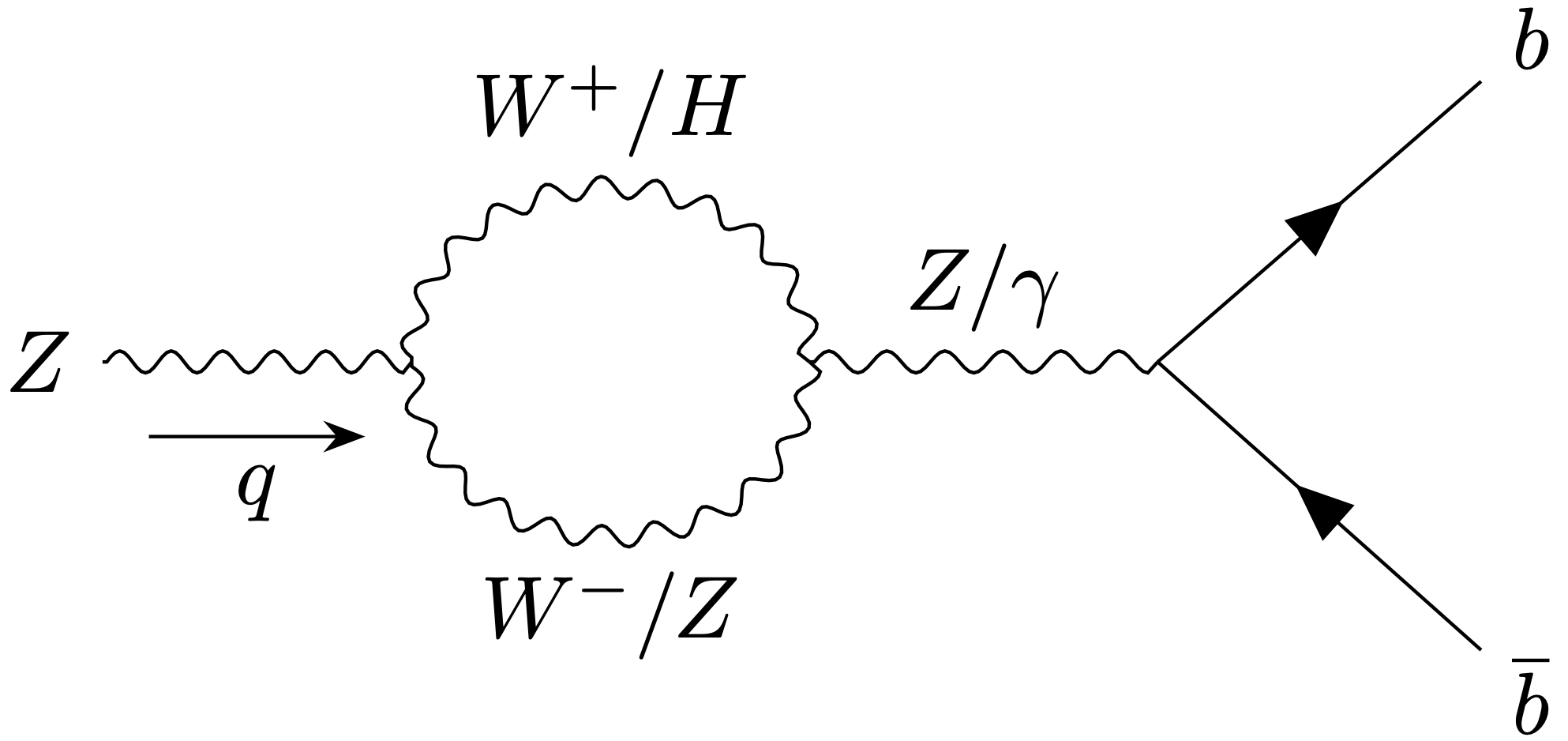}
      \caption{}	\label{fig:Z2bbP}
  \end{subfigure}%
\caption{\textit{Diagrams contributing to the $Z \rightarrow b\bar{b}$ decay corresponding to (a) vertex correction and (b) wavefunction renormalisation of $Z$-boson propagator. }}\label{fig:Z2bb}
\end{figure}

The presence of the anomalous gauge couplings also leads to a
modification in the electroweak precision variables, whether these be
the oblique corrections or the fermion-gauge couplings.  Of particular
importance is the $Z b \bar{b}$ coupling and the $\rho$ (equivalently,
$T$) parameter. The effective $Z b \bar{b}$ vertex may be parametrized as
\begin{equation} \label{eqn:z1}
\begin{split}
\mathcal{L}_{Z b \bar{b}} = \dfrac{e}{s_\theta c_\theta} \left [ \left ( g_L^b +\delta g_L^b \right ) \bar{b}_L \hspace{0.25mm} \slashed{Z} \hspace{0.25mm} b_L +  \left(g_R^b + \delta g_R^b\right) \hspace{0.5mm} \bar{b}_R \hspace{0.25mm} \slashed{Z} \hspace{0.25mm} b_R  \right ]
\end{split}
\end{equation} 
where $g_{L}= \left( -1/2+ s_\theta^2/3 \right)$ and
$g_{R}=\left(s_\theta^2/3 \right)$ are the SM values of the
couplings. The one-loop contributions of the dimension-6 operators are
encapsulated by the two diagrams\footnote{There is, of course, a
    diagram analogous to Fig.~\ref{fig:Z2bb}$(a$), but incorporating
    the $ZZH$ vertex instead of the $ZWW$ one. However, being
    suppressed by an extra factor of $m_b^2/m_W^2$, the corresponding
    contribution is negligible in size.} of
Fig.~\ref{fig:Z2bb}. It should be noted here that, unlike in the
  previous two cases, an analysis with eqn.(\ref{eqn:LEP}) is no
  longer appropriate for it does not possess the full $SU(2) \otimes
  U(1)$ gauge invariance. With a treatment of the oblique corrections
  being contingent on this gauge invariance, the use of full
  eqn.(\ref{eqn:EFT}) becomes mandatory.
With the most sensitive data on  $\delta g_{L,R}^b$ being
obtained at the $Z$-pole, it is convenient to separate the
corrections  into two parts. One of these originates from the
wave-function renomalization of the $Z$ boson due to a one-loop
oblique correction $\Pi_{ZZ}$ (Fig.~\ref{fig:Z2bbP}) and is given by\footnote{It should be remembered that, in the spirit of EFTs, only terms linear in the
  Wilson coefficients should be retained.} 
%%%
\beq \label{eqn:z3}
\barr{rcl}
(\delta g_{L,R}^b)_{ob} & = & \dis \frac{- \alpha_{em}}{4\pi \Lambda^2} g_{L,R} {\cal A}
\\[2ex]
{\cal A} & \equiv & \dis
\frac{g v\, m_Z}{4c_\theta } 
     \Bigg(\frac{m_H^2}{2 m_Z^2}+\frac{2}{3}\Bigg)
     \Bigg(\frac{ c_B}{c_\theta^2}
          +\frac{c_W}{s_\theta^2}\Bigg) \log\frac{\Lambda ^2}{m_H^2}
     \\[3ex]
     & + & \dis \frac{m_Z^2}{2 s_\theta^2 }\,
     \log \frac{\Lambda^2}{m_W^2}
     \\[2ex]
     &\times& \dis
     \Bigg( 3 (1 - 6 c_\theta^2) g^2 \frac{m_W^2}{m_Z^2} c_{WWW}
     + \left[ 4 c_\theta^2 - \frac{5}{6} \right] c_W
     \\[2ex]
     &&\hspace{1em} \dis 
     -  \frac{1 + 4 c_\theta^2 - 36 c_\theta^4}{12 c_\theta^2}
          \, \left[c_\theta^2  {c_W} - s_\theta^2 c_B \right]
      \Bigg)
\earr
\eeq
The second contribution emanates from the direct one-loop correction
to the vertex (Fig~\ref{fig:Z2bbV}). Applicable only to
the left-handed
coupling\footnote{There is, indeed, a vertex correction engendered by the
  anomalous $ZZH$ coupling that contributes to $\delta g_R^b$. However,
 the contribution is suppressed by the bottom mass.}, it is given
by 
%%%
\beq \label{eqn:z2}
\barr{rcl}
(\delta g_L^b)_{v} &=& \dis
\left[ \frac{\alpha_{em}\,\, \vert V_{tb}\vert^2}{16\, \pi\, s_\theta^2}\;
    \frac{m_Z^2}{2\Lambda^2} \right] \, {\cal B} \,
\log \frac{\Lambda^2}{m_W^2}
\\[3ex]
{\cal B} & \equiv & \dis 
\left[c_W-s^2_\theta \left( c_W + c_B \right)\right]
\left(-1 + \frac{x}{2\,} - \frac{1}{6}\, \frac{m_Z^2}{m_W^2}\right)\, \\[3ex]
	& -& \dis c_W  c^2_\theta\,  \Bigg( - \frac{5m_Z^2}{3\, m_W^2}\, +\, 3 x \Bigg)  ,
\earr
\eeq

Tracing the divergences, we again
note that, unlike in the preceding
cases, the GIM mechanism is not operative here, resulting in quadratic
and higher divergences as well. However, following the arguments in
refs. \cite{PhysRevLett.69.3428,Burgess:1992gx}, we presume that such
loop contributions provide the correct dependence on the new physics
scale only upto terms scaling logarithmically with $\Lambda$ and that
higher divergences would be cancelled by counter-terms in the EFT. The
expressions in eq.(\ref{eqn:z2}) differ from those given in
ref.~\cite{Bobeth:2015zqa} by a relative sign between the $\Delta
\kappa_Z$ and $\Delta g^Z_1$ contributions.  The oblique correction
$\Pi_{ZZ}$ in eq. (\ref{eqn:z3}) matches with that calculated in
ref~\cite{Burgess:1993qk} for the $WW$-loop. Besides, we have also
calculated the $ZH$-loop contribution to the $Z$-propagator which
hasn't been included in ref.~\cite{Bobeth:2015zqa}.

The logarithms in above equations are associated to the RG evolution
between $\Lambda$ and $m_W$. 
As argued in
ref.~\cite{Bobeth:2015zqa}, since
$y_t^2/(4\pi)^2\, \ln(\Lambda^2/m_W^2) \ll 1$
(with $y_t$ being the top-quark Yukawa coupling) resumming the corresponding leading-logarithms
  is not numerically important and retaining only the unresummed
  $\ln(\Lambda^2/m_W^2)$ contribution, as above, is a very safe approximation.

\section{Results}

TGCs have been probed in many collider experiments over the years and
deviations from the SM values have been increasingly constrained with
an increase in the collision energy.  The most recent studies by ATLAS
and CMS experiments at the LHC
\cite{CMS:2017egm,ATLAS:2016zwm,CMS:2015tmu,CMS:2019ppl,CMS:2020mxy,ATLAS:2019rob}
have primarily analysed multiple production channels such as $W^+
W^-$, $W^\pm \gamma$ and $W^\pm Z$ with the first-mentioned proving to
be the most restrictive. Each such underlying process can result in
different final states and several have been explored. Of the two most
sensitive studies, one~\cite{CMS:2020mxy} considers the $W^+W^-$
channel with both $W$s decaying leptonically, leading to a $l_a^+ \nu_a l_b^-
\bar{\nu}_b$ final state. The major reducible backgrounds emanate from
the Drell-Yan production of lepton pairs and $t \bar{t}$ events in
which both top quarks decay leptonically.  The Drell-Yan events can be
suppressed by selecting the two leptons to be of different flavor (one
electron and one muon), wheras contributions from $t \bar{t}$ events
can be reduced by rejecting events with b-tagged jets. The second
important analysis~\cite{CMS:2019ppl} considers both $W^+ W^-$ and
$W^\pm Z$ production where one $W$ decays leptonically and the other
$W/Z$ decays hadronically resulting in a $l^\pm \nu q \bar{q'}$ final
state.  In this case, the two major backgrounds are $W + jets$ and $t
\bar{t}$, and a major portion can be reduced by using a combination of
kinematical cuts and jet substructure techniques.  On the other hand,
LEP experiments also posit bounds on the same operators from $W^+W^-$
and single $W$ production channels~\cite{ALEPH:2013dgf}.  Of the
several studies, the most stringent bounds on the anomalous TGCs, in
terms of the Wilson coefficients $c_B/\Lambda^2$, $c_W/\Lambda^2$ and
$c_{WWW}/\Lambda^2$, have been provided in \cite{CMS:2019ppl} for the
LHC operating $\sqrt{s}=13$ TeV and we use these for all subsequent
comparisons. We note that the one-parameter best fit limits on
$c_B/\Lambda^2$ and $c_W/\Lambda^2$ from Higgs production and decay
measurements presented in ref. \cite{CMS-PAS-HIG-19-005}, although
stricter, are comparable with those from the TGC
measurements. However, we do not refer to these limits in our analysis
explicitly as the ensuing sections (\ref{subsec:bands}--\ref{subsec:proj})
deal only with two-dimensional
parameter spaces in the WCs.  In what follows, we compare the current
limits from the LHC against those obtained from the aforementioned
observables for a benchmark value of the cutoff scale $\Lambda$.

A particular point needs to be noted at this juncture. An EFT such as
that in eqn.(\ref{eqn:EFT}) should, ideally, be used in calculating cross
sections only when $\sqrt{\hat{s}} \lesssim \Lambda$, where, $\hat{s}$ denotes the partonic center-of-mass energy.  In the case of the LHC,
this would nominally imply that $\Lambda \gtap {\cal O}(1~{\rm
  TeV})$---even accounting for the nontrivial parton distribution
functions---a constraint that is often overlooked in interpreting
results, generally presented in terms of $c_i / \Lambda^2$. 

In Table~\ref{tab:1} we list the current limits for the aforementioned
observables. For the case of $(g-2)_\mu$ we quote two different values
for the discrepancy $\Delta a_\mu^{\rm anom.}$, namely, $\Delta
a_{\mu}^{\mbox{\tiny DISP}}$ (WP20) and $\Delta a_{\mu}^{\mbox{\tiny
    BMW}}$ (BMW). The limits on $\Delta C_9$ and $\Delta C_{10}$ are
derived from single parameter fits to all $b \to s l^+ l^-$ branching
ratio and angular observables (including that of $\Lambda_b \to
\Lambda \mu^+ \mu^-$) excluding\footnote{The global fit that we refer
  to, from \citep{Hurth:2021nsi}, does not include the $B_{s,d}\to
  \mu^+ \mu^-$ branching ratios. Including this, however, would only
  have a marginal effect on the fit result and, hence, can be safely
  ignored.} the observables sensitive to lepton flavour
  universality (LFU) such as the ratios
$R_{K^{(*)}}$~\cite{Hurth:2021nsi}. Similar fits have also been
performed by other groups, see
e.g.\cite{Carvunis:2021jga,Altmannshofer:2021qrr,Alguero:2021anc}.  On the other hand,
the limits on $\Delta C_7$ are extracted from the inclusive B-meson
radiative decay $(\mathcal{B}_{s \gamma}: B/\bar{B} \to X_s \gamma)$ observable $R_{X_{s}}\equiv \mathcal{B}_{s \gamma}/(\mathcal{B}_{s
  \gamma})_{SM}$ \cite{Altmannshofer:2014rta,Arbey:2018ics}. It is important to note that the
exclusion of the observables such as $R_{K^{(*)}}$, is enforced by
the fact that whereas
their current measurements hint towards a violation of
LFU, the operators under discussion are strictly flavour blind.

\begin{table}[!h]
\begin{center} 
\begin{tabular}{|c|c|c|c|c|}
\hline
\multicolumn{2}{|c|}{Current limits}\\
\hline
Observable($\mathcal{F}$) & $1 \sigma$ limit \\
\hline
\hline
$\Delta a_\mu^{\mbox{\tiny DISP}}$(WP20) & $251 \pm 59 \times 10^{-11}$ \\
\hline
$\Delta a_\mu^{\mbox{\tiny BMW}}$(BMW) & $107 \pm 69 \times 10^{-11}$ \\
\hline
$\Delta C_7$ & $-0.03 \pm 0.03$\\
\hline
$\Delta C_{9\mu\mu}$ &$-1.03 \pm 0.13$\\
\hline
$\Delta C_{10\mu \mu}$ & $0.41 \pm 0.23$\\
\hline
$\Delta C_{9 ee}$ &$0.70 \pm 0.60$\\
\hline
$\Delta C_{10 ee}$ & $-0.50 \pm 0.50$\\
\hline
$\delta g_L$ & $0.0016 \pm 0.0015 $\\
\hline

$\delta g_R$ &$0.019 \pm 0.007 $\\
\hline							

\end{tabular}
\caption{\label{tab:1} {\it Current experimental limits on various observables affected by anomalous TGCs\cite{Hurth:2021nsi,Ciuchini:2013pca}.}}
\end{center}
\end{table}

\subsection{$2\sigma$ bands for all observables for current limits} \label{subsec:bands}

\begin{figure}[htb]  \label{fig:2_cB_cW}
	\centering
\hspace{-5cm}
  \begin{subfigure}{0.2\textwidth}
      \centering
	\includegraphics[scale=0.37,keepaspectratio=true]{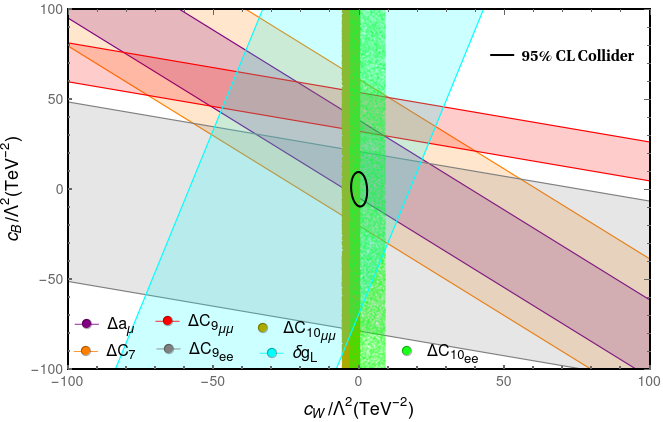}
     \caption{}	\label{fig:2_cB_cW_BMW}
  \end{subfigure}%
  \\ \hspace{-5cm}
  \begin{subfigure}{0.2\textwidth}
      \centering
	\includegraphics[scale=0.37,keepaspectratio=true]{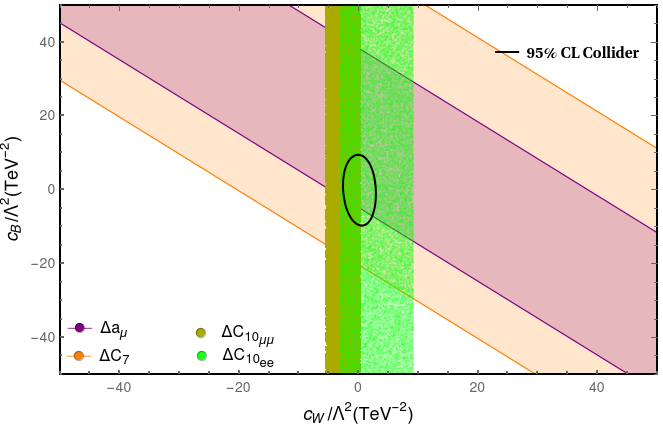}
     \caption{}	\label{fig:2_cB_cW_BMW_zoom}
  \end{subfigure}%
  \caption{\textit{The 2$\sigma$ ranges in the $c_W/\Lambda^2$--$c_B/\Lambda^2$ plane
      allowed by individual observables. $\Lambda = 2$~TeV has been
      used along with the lattice result~\cite{Borsanyi:2020mff}. The
      lower panel shows an enlarged portion of the same. The black
      ellipse identifies the LHC limit.}}
\label{fig:Result1}
\end{figure}

In our quest to study the low-energy constraints in their totality, we
begin by studing each in isolation. Assuming, for the purpose of easy
visualization, that only the Wilson coefficients $c_B$, $c_W$ are
nonzero, we present, in Fig.~\ref{fig:Result1}, the ensuing bounds in this plane emanating from the
individual observables in Table~\ref{tab:1} with the assumption that
the new physics scale\footnote{Note that, unlike the collider bounds
  (which have been derived by neglecting all subleading dependence on
  $\Lambda$), the SMEFT constraints in Fig.~\ref{fig:Result1} have
  additional logarithmic dependence of $\Lambda$, and we would return
  to this point later.}  $\Lambda \sim 2$ TeV.
Several features are worth noting:
\begin{itemize}
\item The $\Delta a_\mu$-allowed band (purple), as calculated using
  the WP20 result~\cite{Aoyama:2020ynm} does not include the SM point,
  reflecting the fact that the data does not agree with the SM value
  at the $2\sigma$ level. For the lattice result
  (BMW)~\cite{Borsanyi:2020mff}, though, it is indeed
  included. For brevity's sake, here,
  and in subsequent discussions, we display only the figures
  corresponding to the BMW result, while keeping under consideration
  the WP20 alternative as well.
  
\item The corresponding band for $\Delta a_e$ sits on the opposite
  side of the origin, owing to the sign of the discrepancy. However,
  with the anomalous contribution being suppressed by $m_e^2$, the
  required sizes of the Wilson coefficients are too large to be
  meaningful.

\item $\delta g_L$ has a relatively weaker dependence on $c_B$ than on $c_W$
  (see eqns.(\ref{eqn:z3}\&\ref{eqn:z2})) leading to the slightly tilted
  band. $\delta g_R$, on the other hand, receives a small correction only from
  the correction to the $Z$ self-energy, and the ensuing bounds are too weak to be relevant.

\item Since  $\Delta C_{7}$, just like $\Delta a_\mu$, parametrizes the coupling
  of a fermion current to the photon, both are understandably proportional
  (in the absence of a nonzero $c_{WWW}$) to the combination $(c_B + c_W)$  and
  the ensuing bands are parallel to each other.

\item $\Delta C_{10}$, being dependent on $c_W$ alone, leads to a
  relatively narrow vertical band in this plane. Most restrictive of
  all the observables, the difference in its value as calculated from
  the electronic and muonic channels exert opposite pulls leading to
  the two parallel bands.  Although both bands overlap with the
  collider limit, the one corresponding to the muonic channel has a
  greater sensitivity to $c_W$ and, hence, its partial overlap
  presents a comparatively stronger constraint on the allowed region,
  favouring negative values for $c_W$.

  This leads to an interesting possibility wherein $\Delta C_{10}$ is
  the dominant flavour-blind Wilson coefficient parametrizing new
  physics effects in FCNC $B$ decays.  A sizable range of $c_W$ values
  compatible with the LHC limits exists that could, then, ameliorate
  the discrepancies in the aforementioned $B$ decay observables
  (excluding LFU ones).  Not contributing to $\Delta C_{10}$, a similar-sized
  $c_B$ would lead to only tiny changes in the low-energy observables
  (see the zoomed-in view of Fig.\ref{fig:2_cB_cW_BMW_zoom}) and would
  be primarily constrained by collider experiments.

\item Similar to the preceding observable, the opposing
  experimental numbers for $\Delta C_9$ from
  the two ($e$ and $\mu$) channels lead to two bands. 
  
\item It is quite apparent that the flavor observables $\Delta
  C_{7,9,10}$ give some of the strongest constraints. Indeed, with
  many of the bands intersecting each other at different angles, it is
  expected that combining the individual data, when independent, would
  lead to constraints much stronger than individual ones. Whether such combined constraints agree with
  the LHC results is an aspect which we address in the following
  subsection.
\end{itemize}

\begin{figure}[htb]  %\label{fig:2_cBcW_cWWW}
	\centering
\hspace{-5cm}
  \begin{subfigure}{0.2\textwidth}
      \centering
	\includegraphics[scale=0.36,keepaspectratio=true]{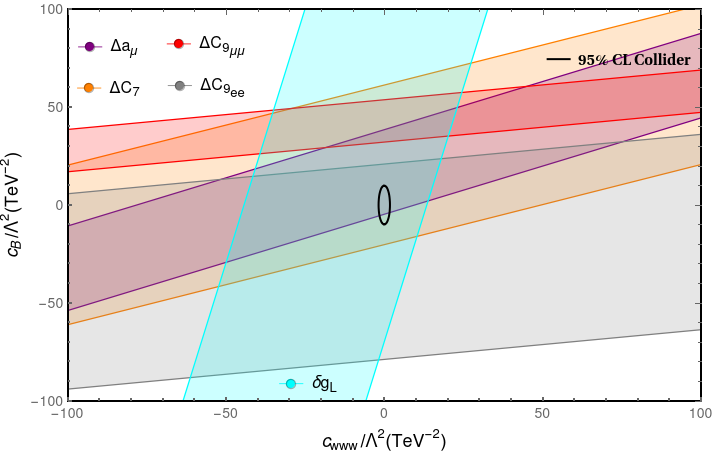}
\hspace{3.5cm}         
      \caption{}	%\label{fig:2_cW_cWWW_BMW}
  \end{subfigure}%
\\
\hspace{-5cm}
  \begin{subfigure}{0.2\textwidth}
      \centering
	\includegraphics[scale=0.36,keepaspectratio=true]{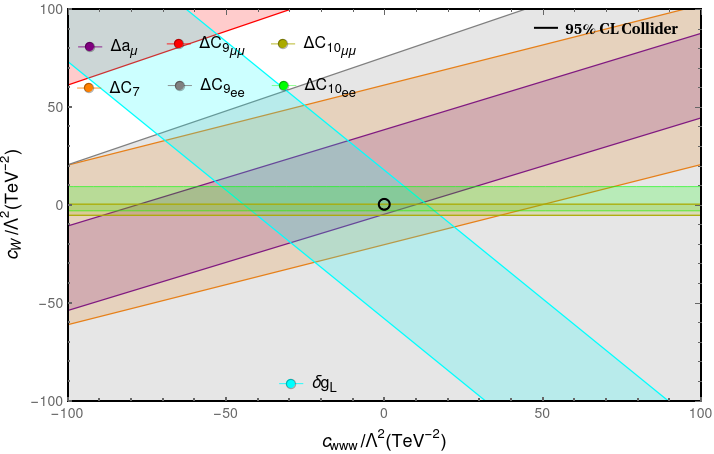}
      \caption{}	%\label{fig:2_cB_cWWW_BMW}
  \end{subfigure}%
	\caption{\textit{The 2$\sigma$ bands as derived from the
            individual observables from anomalous TGCs considering BMW
            result~\cite{Aoyama:2020ynm} as the theory value in the
            $c_{WWW}/\Lambda^2$- $c_{B}/\Lambda^2$ (upper panel) and
            $c_{WWW}/\Lambda^2$- $c_{W}/\Lambda^2$ (lower panel)
            planes along with the 13 TeV experimental
            constraints.}} \label{fig:Result2}
\end{figure}

Before we attempt this, it behoves us to consider the limits
afforded on $c_{WWW}$ and we display these in
Fig.\ref{fig:Result2}, once in the $c_B$--$c_{WWW}$ plane
(keeping $c_W = 0$) and once in the $c_W$--$c_{WWW}$ plane (keeping
$c_B = 0$). With their dependences on $c_{WWW}$ being different, the bands
due to $\Delta a_\mu$ and $\Delta C_7$ are no longer parallel. $\Delta C_9$
continues to be a restrictive force in the $c_B$--$c_{WWW}$ plane, while
its effect is reduced in the  $c_W$--$c_{WWW}$ plane (owing to the smaller
dependence on $c_B$). For $\delta g_L$, it is, quite understandably, the
other way around. 

It needs to be appreciated at this point that the nominal restrictions
from the low-energy observables are much weaker than those obtained at
the LHC, as exhibited by all the panels of
Figs.\ref{fig:Result1}\&\ref{fig:Result2}. This situation changes
drastically once we attempt to simultneously fit all the observables
as we see next. Also worth noting is the fact that much of the
parameter space satisfying the individual observables fall outside the
domain wherein the EFT series expansion can be safely
deemed valid and, hence,
caution must be exercised while reading off such regions as
constraints on the WCs.

\subsection{Fitting all observables} \label{subsec:fit}
We now perform a combined fit of the non-standard parameters $(c_B,\,
c_W, \, c_{WWW})$ to the observables defined in
sec.(\ref{sec:allobs}).  To this end, we define a $\chi^2$ as a
function of the anomalous
parameters, namely
%%%
\beq
\chi^2(c_B, c_W,c_{WWW}; \Lambda)= \sum_i \left( \frac{\mathcal{F}^{\rm exp}_i-\mathcal{F}^{\rm th}_i}{\sigma_i}\right)^2,
\label{chi_sq}
\eeq 
%%%
where $i$ runs through all the observables, $\mathcal{F}^{\rm exp}_i$
are the experimental values with standard deviations $\sigma_i$ and
$\mathcal{F}^{\rm th}_i$ denote the corresponding theoretical
expectations for a given set of values for the Wilson
coefficients. The best-fit point would then be given by the minimum of
the $\chi^2$ and parameter points leading to $\chi^2 \leq \chi^2_{min}
+ \delta \chi^2$ being inseparable from the best-fit point at a
confidence level determined by $\delta\chi^2$. As before, in order to
render the constraints more tractable, we consider only two nonzero
Wilson coefficients at a time, holding $\Lambda$ to a fixed value.
%%%
\begin{figure*}[htb] 
	\centering
\begin{subfigure}{0.2\textwidth}
      \centering
      \hspace{-6cm}
	\includegraphics[scale=0.35,keepaspectratio=true]{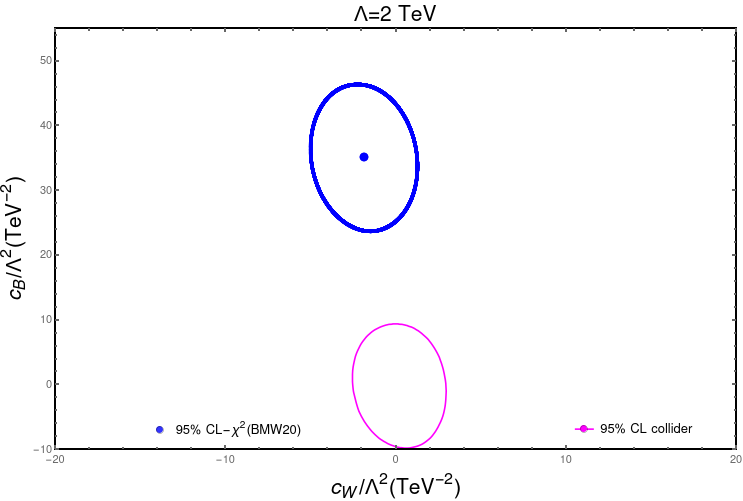}
	\caption{}  \label{fig:cb_cw_current}
  \end{subfigure}%
%\hspace{6cm}  
\begin{subfigure}{0.2\textwidth}
      \centering
      \hspace{3cm}
	\includegraphics[scale=0.35,keepaspectratio=true]{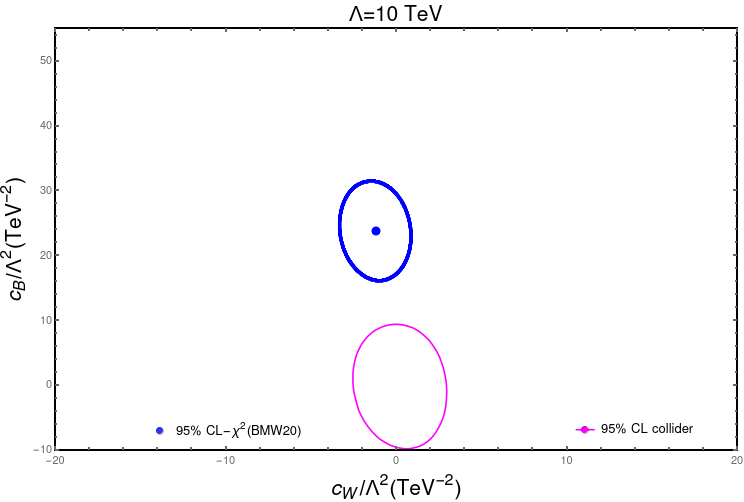}
	\caption{}  \label{fig:cb_cw_current_10TeV}
  \end{subfigure}%  
\caption{\textit{\ref{fig:cb_cw_current} shows the result of the $\chi^2$ analysis in the BMW case for $\Lambda=2$ TeV, and the collider result at $95 \%$ confidence level, in the
$c_W/\Lambda^2$- $c_{B}/\Lambda^2$ plane. \ref{fig:cb_cw_current_10TeV} shows the same for $\Lambda=10$ TeV.}} \label{fig:Result3}
\end{figure*}

The calculation of the $\chi^2$ entails inclusion of either only
independent measurements, or if more variables are to be considered,
an appropriate inclusion of the correlation matrix. For the sake of
simplicity, we adopt the former method despite the attendant loss of
sensitivity. This allows us to use the corresponding measurements
(alongwith the errorbars) from Table \ref{tab:1} in a straightforward
manner. In particular, thanks to the flavour universal nature of the
  Wilson coefficients in the theory, we make use of both the muonic
  ($\Delta C_{\mu \mu}$) as well as the electronic ($\Delta C_{ee}$)
  limits simultaneously.

\begin{table}[!htb]
\begin{center}
\begin{tabular}{|c|c|c|c|}
\hline
Calculation & Descriptor & $(c_B, c_W, c_{WWW})/\Lambda^2 $ & $\chi^2$\\
& & [TeV$^{-2}$] & \\
\hline
\hline
WP20  & SM & (0,0,0) & 101.76\\
& 2-param B.F. & $(39.26, -1.64, 0)$ & 25.76\\
& 3-param B.F. & $(38.48, -1.63, -2.97)$ & 25.71\\
\hline
\hline
BMW  & SM & (0,0,0) & 86.121\\
& 2-param B.F. & $(35.05, -1.83, 0)$ & 28.663\\
& 3-param B.F. & $(36.65, -1.85, 6.41)$ & 28.446 \\
\hline
\hline
\end{tabular}
\end{center}
\caption{\it Results of the different $\chi^2$ analyses described in the text.}
\label{tab:chisq}
\end{table}

We, thus, have two inequivalent scenarios (depending on whether we
choose the WP20 or BMW calculation) that the theory is to be compared with. The
resultant $\chi^2_{min}$ values are
tabulated in Table~\ref{tab:chisq}. Also tabulated, for comparison, are the $\chi^2$ values for the SM point. The strikingly different $\chi^2$ values for the latter under the two schemes is but a consequence of the
  difference in the theoretical predictions for $(g-2)_\mu$. Also,
note that the rather large values
for the SM point are driven, to a large extent, by
the flavour universal $B$-decay observables.

To interpret these results, it is instructive to consider a projection
to a two-parameter space, which we consider to be the ($c_W/\Lambda^2,
c_B/\Lambda^2$) plane with $c_{WWW}$ held zero.  The corresponding
global fit, in the best case scenario ({\em i.e.}, BMW), is illustrated in
Fig.\ref{fig:cb_cw_current} wherein we show the best-fit with the
attendant 95\% C.L. contour.  We note that the $\chi^2$-fit not only
prefers the best-fit value to be far away (as determined on the scale
of the LHC limits) from the SM value, but even the two 95\% ellipses
(direct limits and low energy preference) do not overlap with each
other.  That the separation is larger along the $c_B$--axis can be
understood qualitatively by comparing the individual bands in
Fig.\ref{fig:2_cB_cW_BMW}. On account of the relatively
small uncertainty associated with it, the muonic $\Delta C_9$
  measurement exerts the strongest pull on $c_B$, thereby pulling the
best-fit point upward.

As can be gleaned from eq.\ref{eq:DeltaC710calculation}, and as has
already been reflected in Fig.\ref{fig:Result2}, the observables under
consideration have only a very weak dependence on
$c_{WWW}$. Consequently, its contribution to the $\chi^2$ is
relatively small, and, on its inclusion in the minimization, the
erstwhile minimum in the two-parameter fit (with $c_{WWW} = 0$) only
expands into a very shallow basin. This is reflected by
Table~\ref{tab:chisq}, where the inclusion of
 $c_{WWW}$ doesn't have any significant impact on the $(c_B, c_W)$
coordinates of the best-fit point. Similarly, the improvements in the
 attendant $\chi^2_{\rm min}$ values are only marginal.

A caveat needs to be discussed at this point. Although we have been
characterizing our fits as functions of $c_i/\Lambda^2$, in actuality
we have held $\Lambda$ to a specific value (2 TeV) especially in the
calculation of the logarithms. Increasing it to higher values not only brings the best-fit point closer to the SM, but also, understandably, shrinks the 95\% C.L. ellipse significantly (see
  Fig.~\ref{fig:cb_cw_current_10TeV}). The apparent tension with the
  LHC constraints is maintained. This disagreement is expected to
  persist as, with additional luminosity (LHC Run III, HL-LHC, etc.)
or higher energies (a future collider or an upgrade of the LHC), the
collider limits are likely to shrink too (as it already has for the 13
TeV run as compared to the 8 TeV one\cite{CMS:2019ppl}).

%\begin{figure}[h!] 
%\centering
%\includegraphics[scale=0.3]{figures/electron_current_10TeV_WP_BMW_combined.png}
%\caption{\textit{The $\chi^2$-fit in the electronic scenario for the benchmark %case of $\Lambda=10$ TeV in the $c_W/\Lambda^2$- $c_{B}/\Lambda^2$ plane.}} \label{fig:e_10tev}. 
%\end{figure}

  %%%%%%%%%%%%%%%%%%%%%%%%%%%%%%%%%%%%%%%%%%%%%%%%%%%%%%%%%%%%%%%%%%%

%%%%%%%%%%%%%%%%%%%%%%%%%%%%%%%%%%%%%%%%%%%%%%%%%%%%%%%%%%%%%%%%%%%
%\newpage
\subsection{Future Projections} \label{subsec:proj}
The analysis in the preceding subsection establishes that an effort to
explain $(g-2)_\mu$ or several other discrepancies in the low-energy
data in terms of anomalous triple-gauge boson couplings (or,
equivalently, the corresponding bosonic operators in the SMEFT) would
require couplings (Wilson coefficients) that are too large given the
LHC constraints. It is, then, of interest to 
speculate whether near-future improvements in the low-energy
data are likely to lead to constraints stronger than those already
imposed by the LHC.  As a particular example, we choose to do
  this for $(g-2)_\mu$. With the experiment at
FNAL projecting a reduction in the uncertainties by a factor of about
four \cite{Aoyama:2020ynm}, it is worth re-examining our analysis with
the following two assumptions as to the tentative outcomes by the end
of the FNAL experiment(s):

\subsubsection{Reduced uncertainties in $\Delta a_\mu$ with the same central value}
\begin{figure}[h!] 
\centering
\includegraphics[scale=0.35]{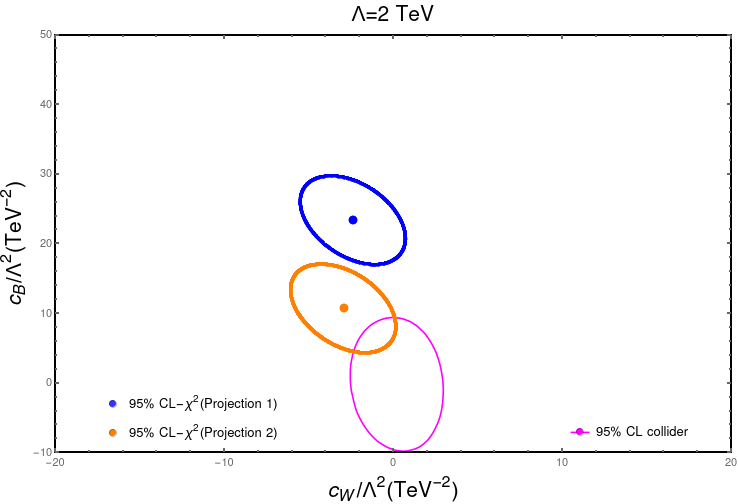}
\caption{\textit{The $\chi^2$-fit and the collider result at $95 \%$ confidence level for $\Lambda=2$ TeV in the $c_W/\Lambda^2$- $c_{B}/\Lambda^2$ plane: (a) (blue) assuming same deviations as of BMW result and corresponding errors reduced by a factor of 4 and (b) (orange) assuming no deviations from the SM result and errors reduced by a factor of 4.}} \label{fig:future_proj} 
\end{figure}

With no significant improvement in the theoretical calculations
  envisaged in the near future, if the central value of the experimental measurement
  remains unchanged, we would be likely to face $\Delta
a_\mu^{\mbox{\tiny DISP}}=(251 \pm 15)\times 10^{-11}$ ($\Delta
a_\mu^{\mbox{\tiny BMW}}=(107 \pm 17.15)\times 10^{-11}$) pertaining
to the WP20 (BMW) analyses. The
corresponding $\chi^2$-fit for the projected $\Delta
a_\mu^{\mbox{\tiny BMW}}$ (best case scenario) is shown in
Fig.\ref{fig:future_proj}(blue curve). Comparing with the current limits presented
in Fig.\ref{fig:cb_cw_current} we note that while the best fit point
has shifted marginally towards the SM point along the $c_B/\Lambda^2$
direction, the corresponding $95 \%$ C.L. contour has, expectedly, shrunk. The overall tension would actually increase.

\subsubsection{Reduced uncertainties in $\Delta a_\mu$ with a vanishing central value}
In the other hypothetical case of a vanishing discrepancy by the end
of the FNAL run, we would have $\Delta a_\mu=0 \pm (17.15)\times
10^{-11}$, and Fig.\ref{fig:future_proj}(orange curve) shows the
corresponding global fit. The best-fit point would now shift by a
significant amount towards the SM point with the contour ellipse
having shrunk as in the preceding case.  Furthermore, the best-fit ellipse now has a small overlap with the LHC contour. This fitting, thus, gives a
true measure of the pull exerted by the cohort of low-energy
flavour-universal observables (other than $\Delta a_\mu$).
  
 Note that the preceding projections correspond specifically to $\Lambda=2$ TeV. For higher cutoff scales, the best fits would shift considerably towards
 the SM point, as had been indicated in
 Fig. \ref{fig:cb_cw_current_10TeV}, and would  offer
 a better reconciliation with
 the LHC limits as well. However, any claim about this indicating a
 resolution of the discrepancies through the WCs residing in the
 overlap are fraught with danger. For one, were hypothetical
 resonances that generate these WCs interacting only weakly with each
 other and with the SM sector, one would expect $\mathcal{O}(g^2_{SM}/16 \pi^2)\lesssim(|c_B|, |c_W|)\lesssim
 \mathcal{O}(1)$~\cite{Marzocca:2020jze,Giudice:2007fh}, with $g_{SM}$
 denoting a typical SM coupling.  However, much of the overlap region
 would most probably correspond to $(|c_B|, |c_W|) > \mathcal{O}(1)$
 indicating a strongly interacting sector that calls into question the
 very method of calculating quantum corrections that we have
 adopted. In particular, no definite conclusion with regards to the
 matching of the SMEFT operators with a UV theory can be drawn.

\section{\label{sec:Conclusion} Summary and Outlook}

Our study indicates that the limits on low-energy observables, taken
individually, lead to weak bounds on the bosonic SMEFT Wilson
coefficients when compared with the existing LHC limits, except for
the bounds on $c_W/\Lambda^2$ emanating from the limits on $\Delta
C_{10}$ which are comparable and consistent with the collider
results. On the other hand, a global fit in the $(c_W/\Lambda^2,
c_B/\Lambda^2)$ plane, while imposing significantly stronger
constraints on the WCs, exhibits disagreement with the LHC
results. The naive expectation is that this
disagreement would reduce with a
(speculative) lowering of the $a_\mu$ tension in the future; on
  the other hand, improvements in the collider limits on the WCs are as likely to
  maintain the disagreement. At this juncture a cautionary remark
ought to be underscored to aid our judgement. The LHC limits have used
cross-sections that also include terms quadratic in the TGCs, whereas
we consider contributions to the concerned observables only upto a
linear order in the same. Had we included quadratic contributions, we
would have obtained stronger bounds as well. Thus, in view of this
caveat, it would be rash to rule out completely the scenario wherein
anomalous TGCs embody some of the most dominant new physics effects
addressing all of the aforementioned anomalies.

Notwithstanding the caveats, the fact that $\Delta C_{10}$ also favours $c_W/\Lambda^2$ values that are very close to the origin indicates that any explicit new physics model designed to explain the discrepancies (e.g., models which give rise to lepton flavour
  universality violating (LFUV) or a combination of LFUV and LFU
  4-fermion operators\footnote{See ~\cite{London:2021lfn} for a
    review.}) should either induce $\mathcal{O}_W$ with a suppressed (or vanishing) Wilson coefficient or, otherwise, one must account for the WC $c_W$ generated therein, in addition to other parameters, while performing a fit to the concerned observables. This is a crucial outcome of our study, to be regarded as a check on the existing models concerning low-energy observables, as well as a note to the model builders developing new scenarios along this line.
%%%%%%%%%%%%%%%%%%%%%%%%%%%%%%%%%%%%%%%%%%%%%%%%%%%%%%%%%%%%%%%%%%%
%%%%%%%%%%%%%%%%%%%%%%%%%%%%%%%%%%%%%%%%%%%%%%%%%%%%%%%%%%%%%%%%%%%

%\input{sections/conclusion.tex}

%%%%%%%%%%%%%%%%%%%%%%%%%%%%%%%%%%%%%%%%%%%%%%%%%%%%%%%%%%%%%%%%%%%%

%%%%%%%%%%%%%%%%%%%%%%%%%%%%%%%%%%%%%%%%%%%%%%%%%%%%%%%%%%%%%%%%%%

%%%%%%%%%%%%%%%%%%%%%%%%%%%%%%%%%%%%%%%%%%%%%%%%%%%%%%%%%%%%%%%%%%%

\begin{acknowledgments}
We thank Martin Hoferichter and Diego Guadagnoli for insightful
comments and for bringing to our notice some important
references. K.D. acknowledges Council for Scientific and Industrial
Research (CSIR), India for JRF/SRF fellowship with the award letter
no. 09/045(1654)/2019-EMR-I. S.M. acknowledges research Grant
No. CRG/2018/004889 of the SERB, India. L.K.S. acknowledges the UGC
SRF fellowship and research Grant
No. CRG/2018/004889 of the SERB, India, for partial financial support. 
%=================================================================
\end{acknowledgments}

%%%%%%%%%%%%%%%%%%%%%%%%%%%%%%%%%%%%%%%%%%%%%%%%%%%%%%%%%%%%%%%%%%

%\appendix

%\section{Appendix 1}

%%%%%%%%%%%%%%%%%%%%%%%%%%%%%%%%%%%%%%%%%%%%%%%%%%%%%%%%%%%%%%%%%%

\bibliographystyle{apsrev4-1} % Tell bibtex which bibliography style to use
\bibliography{reference.bib} % Tell bibtex which .bib file to use (this one is some example file in TexLive's file tree)

\end{document}